# Radon Decays and Their Implications for Elementary Particle Physics

C. Scarlett[a], E. Fischbach[b], B. Freeman[b], J. J. Coy[c], P. Edwards[d], R. Burkhart[e], O. Piatibratova[f], and T. Monsue[g]

[a] Department of Physics, Florida A&M University

[b] Department of Physics and Astronomy, Purdue University

[c] Ball State University

[d] College of Coastal Georgia

[e] Independent Researcher

[f] Geological Survey of Israel

[g] NASA Goddard Space Flight Center

**ABSTRACT:**

This paper presents a new analysis of the observations of radon decay in an enclosed environment by the Geological Survey of Israel (GSI) between 2007 and 2012 [4] – for a more complete list of experiments performed by GSI on radon see also [1-3]. The data exhibit a large peak around local noon followed by an abrupt drop, and by a second peak around 6PM local time. Additionally, there is also a very low amplitude peak occurring before daybreak. The salient features of the GSI radon decay data can be modeled as arising from a change in the radon decay rate ($\tau$), rather than due to the changes in the local concentration of radon ($N_0$). Such a model may provide a clue to long theorized axionic, dark matter, interactions. Finally, new experimentation is suggested that can distinguish between changes in $N_0$ versus changes in $\tau$. Should a follow-up experiment show an effect similar to that seen in GSI, this could have significant implications for elementary particle physics.

## 1.0 Introduction:

The study of radioactive decay has led to successful models of nuclear structure, explaining phenomena as varied as nuclear stability, decay half-life ($\tau$), and parity violation in beta decays. In heavy nuclei that alpha decay, models describing the radioactive transitions rely on quantum tunneling through the nuclear barrier. Such tunneling is well known to be exponentially dependent on the energy of the released alpha particle, scaling thirty-five orders of magnitude ($10^{+35}$) for energy changes of as little as ~ 4 MeV [5-8]. This makes alpha decay an ultra-sensitive probe of the available excess energy within a nucleus. This also means that energy fluctuations of the outgoing alpha by as little as $10^{+2}$ eV ($10^{-4}$ MeV) can yield a few percent change in decay lifetime. This work looks at whether evidence supports a modification to the calculation of $\tau$ due to external influences that can describe the GSI observations.

Over ten years of data on radon decays have been acquired by the Geological Survey of Israel (GSI). As described by the authors in Ref. 1, the motivation for the GSI experiment was in part to understand the origins of the "large temporal variations" in $^{222}Rn$ decays noted "that differ significantly from variations of (i) other trace elements in geogas (noble gases); (ii) variation patterns of other dynamic geophysical systems." As will become clear from the ensuing discussion of our own data, radon decays are likely to be extremely sensitive to external environmental influences due the creation of $^{222}Rn$ along the "valley of stability" in the Periodic Table. It is the characteristic of $^{222}Rn$ which singularly distinguished its decays from those of other inert gases, such as He, Ne, Ar, Kr, and Xe.

To arrive at their conclusion that "Broad consensus exists that there is no simple and straightforward understanding of the phenomena and behavior of radon…", the authors of Ref. 1-4 carried out a series of laboratory tests in which they sometimes allowed to vary and sometimes controlled environmental parameters such as the local pressure P and temperature T, as well as the concentrations of radon in their various samples. A schematic diagram of one of their setups is shown in Fig. 2 of Ref. 4. As the authors note, "… an elevated level of radon in air is maintained within a tight volume (container) through the diffusion of radon from attached sources". These sources include an industrial source of intensity 103.2 kBq, and a geological source containing uranium with an intensity of 0.5-3kBq. The decay alpha particles were detected by two different instruments: BT45N (Algade Inc., France) and AM-611 (Alphanuclear Inc., Canada). Further details on the instrumentation can be found in Ref. 1.

Since temperature variations are obvious candidates as explanations for the observed variations in the number of detected alpha particles, Steinitz et. al. carried out a series of tests to exclude temperature fluctuations as a possible explanation for the observed time-dependent variations in detected radon decays. These tests included demonstration that two detectors at the same location, but oriented in different directions yielded "…different apparent half-lives during decay runs as well as directionality related global orientation." This observation motivated

Steinitz, et. al. to conclude that "… the location of the artificial activity generating this influence and its nature is unknown."

In what follows, we put forward a model to explain the GSI data based on a proposed axionic contribution from dark matter. The central feature of this model is the recognition that the location of radon in the Periodic Table suggests that the rate of $^{222}$Rn decay will be extremely sensitive to external perturbations, such as might arise from hypothesized axions. Should this model prove to be successful in accounting for various features of $^{222}$Rn decays, then by implication it will also lends support to the hypothesis that axions may be a significant component of dark matter. The specific characteristic of axions which is relevant in the present context is that they are electrically neutral, and relatively weakly interacting. These properties could allow axions to penetrate a nucleus and thereby impart to the $^{222}$Rn nucleus an additional energy that could drive the observed decay fluctuations.

This work looks at the period between 2007-2012 **[1-4]**. During this period, the GSI data show diurnal and annual oscillations that do not appear consistent with variations in background concentrations as suggested elsewhere – a follow up paper shows detailed comparisons of the oscillations to temperature and weather variations, while earlier work by Steinitz and Sturrock have eliminated correlations to power fluctuations. Here it is proposed that these oscillations may be understood in terms of a mechanism known as the Primakoff effect, whereby low energy photons produce and scatter axionic, dark matter particles that in turn can interact with nuclear matter. What is significant about this effect in the present context, is that it may provide an unexpected connection between radon decays and dark matter.

*Note: The GSI collaboration, led by Dr. Steinitz, performed a number of additional experiments to flesh out the source of the observed oscillations. This work discuss only the ten year data set taken on a device located in an external structure and electronically isolated. We have made no arguments about the numerous other experiments performed by GSI – but may revisit some of the other data sets in the future.

## 2.0 Data Analysis:

### *2.1 Data Review*

Figures 1 a-d show a sample of data from the GSI experiment [1-4] taken in 2007 centered on the four annual seasons – February, May, August and November. What can be seen is diurnal behavior with a total of three peaks: 12pm, 6pm, and approximately 12am to 1am. In addition to the three diurnal peaks, there is a striking rise in the amplitude of the peak above background over the course of one year when going from winter, to spring, to summer and finally to fall. The largest peak, occurring at 12pm, shows an average of 5.2% enhancement above background during January, while in June the noon 12pm peak shows an average of 15.9% enhancement above the background. Qualitatively, there is a factor of three in enhancement of the noon peak above background over the course of the year, and this annual cycling is observed in subsequent years (2008 – 2012) as well.

Figure 2a shows ten periods where the GSI data have been summed every fifteen days, (e.g. Period 1 covers February 9th to February 23rd of 2007) to form a single distribution. Figure 2b shows, for comparison, solar irradiance on the ground over the course of a single year. It is clear that the GSI data have a noon peak with an amplitude that tracks with the increase in solar intensity on the ground, over the course of the year. A more detailed comparison of the noon GSI peak, see Figure 3, reveals that the increase in the detected gamma rays, arising from the daughter particle of radon, have a Gaussian distribution. However, beginning around 3pm, the detected number of decays drops at a rate faster than a Gaussian function, giving the peak an asymmetric structure. The rapid drop off in detected decays, between noon and approximately 3pm daily, often results in an overall rate that dips below the baseline. This baseline is defined as the average decay rate using sideband regions of the visible peaks, for time around 12am on one day until 1am the next day (see in Figure 4).

In addition to the 12pm peak, there are peaks centered at roughly the 6pm and 12am-1am hours. These peaks have lower amplitudes and shorter durations as seen in Figures 1-4. The amplitude of the noon peak shows a consistent oscillation which tracks the intensity of sunlight hitting the Earth. Thus the first order effect involves a mechanism that scales proportionately with solar irradiance.

### *2.2 Nuclear Decay Models*

Quantitatively, nuclear decay is described by an exponential function. The exponential law of decay, equation 1, describes the time-dependence of radioactivity samples as derived from just the internal nuclear structure. The parameters needed to calculate a specific count rate are just the number of available radioactive nuclei ($N_0$), local concentration, and the decay half-life ($\tau$ ) – defined as the time needed for half of the available nuclei to decay:

$$N(t) = No \cdot e^{-0.693 \cdot t/\tau} \quad (1)$$

Any observed number of decays during a time period, e.g. per second, can be used to determine: (1) the number of nuclei available to decay, if the half-life is known, or (2) the half-life, provided the number available nuclei is known. The environment outside the nucleus is assumed to not impact nuclei half-life. In theory, once a material's half-life is known, any observed fluctuations in N(t) arise from variance in $N_0$. However, alpha decay is ultra-sensitive to the energy available to the outgoing alpha particle **[5-8]**. Allowing for a half-life that changes as a function of external parameters, which can also be time-dependent, equation (1) can be rewritten:

$$N(t) = No \cdot e^{-0.693 \cdot t / \tau(x(t))} \quad (2)$$

Distinguishing between changes in $N_0$ and $\tau(x(t))$ requires careful experimentation to control and account for the local concentrations $N_0$.

It has always been assumed that the intense energies required for most nuclear excitations mean that the decay half-life of a specific nucleus is in fact constant. Dating back to some of the first radium experiments, Madame Curie et al. attempted to stimulate radioactivity **[9-12]**. Data taken from cooling, heating and pressurizing radioactive materials have shown only infinitesimal changes that have been written off as near the limit of experimentation. However, the Geiger-Nuttall model of heavy radio-isotopes that undergo alpha decay gives an amazingly steep curve for relatively small energy changes. Equation 3 shows how the law derives from the nuclear parameters:

$$\ln \tau \sim 2G + \ln \frac{2R}{v} \quad (3)$$

The second term in equation 3 is the constant parameter, dependent on the nuclear radius (R) and the velocity of the alpha particle in the nucleus. The exponential factor, is called the Gamow Factor (G), and is related to the number of protons (Z), the square-root of the energy of the alpha (E) and an energy parameter (3.9 MeV).

$$G \sim Z'\left(\frac{3.9\ MeV}{\sqrt{E}}\right) \quad (4)$$

Figure 6 (a) shows a graph of decay half-life for Thorium, Uranium, Actinium and Neptunium as a function of the energy of the ejected alpha. The plot depicts half-life, plotted on a logarithmic scale (y-axis), versus the energy of the ejected alpha particle on a linear scale (x-axis). Using a linear approximation for the data between 4 MeV and 6 MeV, gives $\gg 10^{+12}$ change in half-lives. $^{222}Rn$ has a half-life of 3.8 days ($3.28 \times 10^{+4}$ sec) and, based on the data in Figure 5, the energy of the ejected alpha is, as predicted, E ~ 5.59 MeV. If the GSI data were to be accounted for as a change in the energy of the alpha, given that the oscillations are at the scale of 5-15%, the equations above suggest that energy changes of as little as $10^{+2}$-$10^{+3}$ eV (Extreme-UV or Soft X-ray) are required. ***Note:*** the equations above are approximated using the Geiger-Nuttall formula and the predictions scale as much as 56 orders of magnitude, Figure 6 (b), **[10]** for a range of G = 32-45 [$Z'/\sqrt{E}$].

Another approach to connecting axions to nuclear decay involves looking at whether the presence of an axion field could alter the nuclear radius – through axion-nucleon interactions. In their work on modeling nuclear temperature, see **[5]**, S.S. Hosseini et al. show that the nuclear temperature scales exponentially with changes in nuclear radius, see equation 5 and Figure 6 (c):

$$n(T) \sim exp^{[\frac{R(T)-\alpha}{\beta}]} \quad (5)$$

The energy of the ejected alpha scales linearly with the nuclear temperature. Nuclear half-life scales exponentially with alpha energy – Gamow factor becomes $\tau \sim exp^{[Z'(\frac{3.9}{\sqrt{\eta \cdot \exp[\frac{R(T)-\alpha}{\beta}]}})]}$. Where: $\eta$ is the slope of the line relating alpha energy to the nuclear temperature. Thus the relationship between nuclear radius and nuclear half-life scales as an exponential of an exponential. ***Note:*** the original axion model was derived to explain why neutron EDM may always align with its spin. This aligning could become the basis for infinitesimal changes to the nuclear radius.

### *2.3 Axionic Coupling Model*

In experimental particle physics, there is a mechanism that connects photons (solar irradiance) to magnetic fields (e.g. Earth's B) and neutral particles, e.g. axion, capable of penetrating nuclear matter – due to absence of a Coulomb barrier. This mechanism involves a Primakoff coupling **[15-27]**. A photon field can couple in the presence of a magnetic field to a background of axionic dark matter, as depicted in Figure 5.

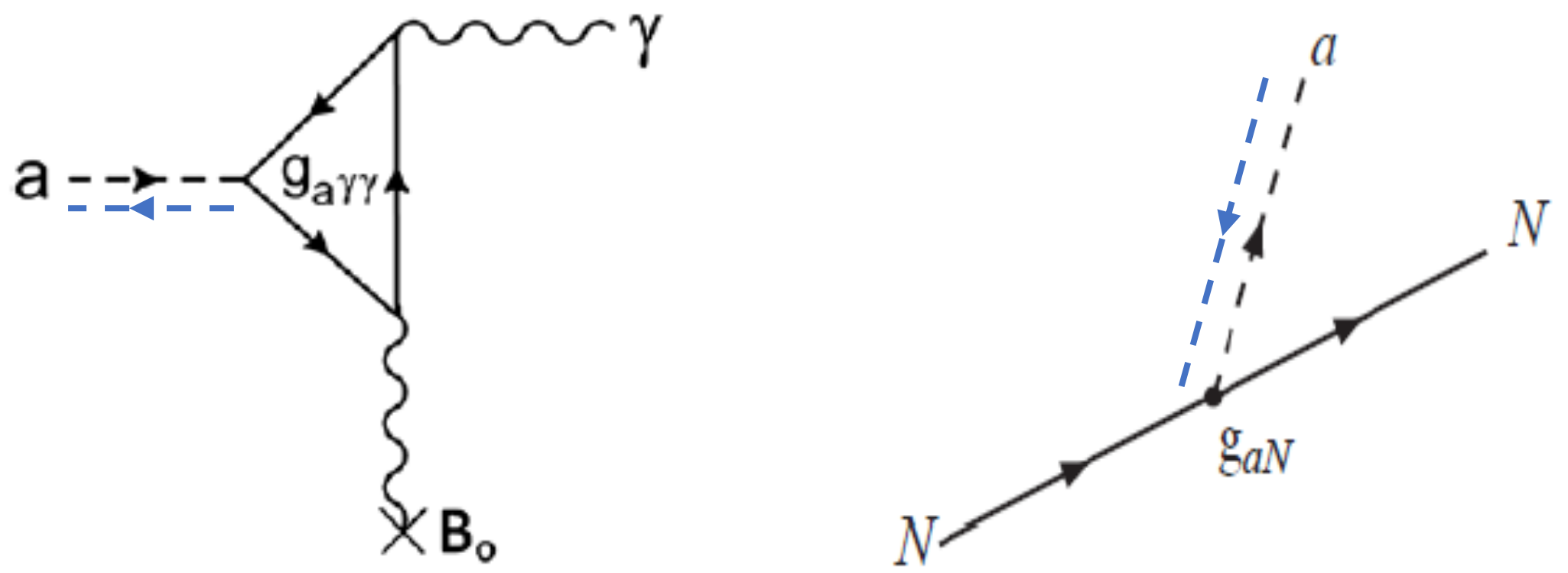

Figure 5: (Left) Inverse Primakoff effect showing connection of a photon, a magnetic field, and an axionic particle. (Right) Primakoff coupling of an axion to a nucleon

The combination of the two diagrams in Figure 5 provides a mechanism whereby photons from the Sun could produce or focus axions, which couple to nucleons in atomic nuclei. Such a mechanism allows the sunlight passing through the Earth's magnetic field to influence axionic matter near the surface of the Earth. The axions, in turn, enable solar energy to influence, even catalyze, nuclear processes on Earth. In fact, axionic dark matter was theorized to couple to quarks to eliminate the neutron EDM. This may explain how changes in the intensity of low energy, solar photons coincide with changes in the rates of decay from certain nuclei[1] – in GSI radon data such a coincidence seems to have been observed. Of course in such a model there is a second, major photon field that must be accounted for – albedo radiation.

Figure 7 shows the two photon fields that bathe the Earth daily. The solar energy peaks at noon for any given region on Earth. The intensity and total hours of daylight depend on the time of year. The Earth reflects radiation, at various positions in the atmosphere and all the way down to the ground, back into outer-space – this is known as albedo radiation **[28-30]**. The peak of surface level (ground) albedo radiation occurs at approximately 3pm – explaining why this is also the hottest part of the day. If a process is impacted by visible energy solar radiation, then, even though the surface level albedo peaks at slightly lower energies, it is reasonable to expect that the same Primakoff process may be impacted by albedo as well. When integrated over the entire day, the albedo radiation will account for much of the solar radiation making it through the

[1] Nuclei have characteristic structure functions which make them undergo distinct types of radioactive decays. For this reason, all nuclei need not necessarily show the same effects, as each will have its own structure function. Hence, susceptibility to interactions with any given type of particle may differ from one nucleus to another

Earth's atmosphere. In a Primakoff model, albedo radiation passing through the Earth's magnetic field will also couple to, or create, axionic dark matter around the Earth.

Figures 8 a-c show three approaches for combining the solar and surface level albedo radiation fields. First, one can do a simple sum as shown in 8a, leading to a Gaussian function. Next one can do a simple subtraction as in 8b, giving an asymmetric function. Finally, the fields can be vector summed as in 8c, accounting for the relative directions of the fields. In panel (a) the amplitude of the total radiation is added together for each 15min period during of the day. The sum of two Gaussian function is a Gaussian function, so the gray curve shown is a single, Gaussian peak at approximately 2pm. In panel (b) the total irradiance due to albedo radiation is subtracted from that of the solar radiation. The graph shows a peak structure that drops below the sideband region (12am on one day and 1am on the next day), only to rise to side band level at approximately 9pm. However, in panel (b) there are only two regions of elevation above the sidebands. One such region occurs three hours after the 6pm peak observed in the actual GSI data. Additionally, the drop below sideband level occurs at 6pm and not 3pm as seen in the GSI data. Panel (c) shows what happens when the solar radiation is added vectorially to the albedo, taking the direction of energy flow into account. In this panel, there are three characteristic peaks at noon, 6pm and between 12am – 1am. Similar to the data, this vector addition in panel (c) displays the characteristic Gaussian rise until about noon, followed by a faster than Gaussian drop off between 3pm and 4pm. The peak at 6pm has approximately the same amplitude relative to the noon peak (see Fig. 3 – GSI sample data 2007), with no need to adjust for the longer wavelengths typical of albedo when compared to solar irradiance.

## 3.0 $^{222}$Rn Decays

Experimental data show that all elements with fewer than 84 protons have at least one stable, non-decaying, isotope. For elements with proton numbers between 90 and 98, at least one or more isotopes is semi-stable or has a lifetime in excess of 750 years – with many isotopes exceeding tens of thousands, millions or even billions of years. Figure 9 presents graphs which summarize our current knowledge of nuclear half-lives and decay modes. Notably, the region for elements with 84-89 protons, known as the "valley of stability," are dominated by both low values for isotope half-lives and a high rate of alpha decays. Near the middle of this "valley of stability" is $^{222}$Rn with 86 protons and 136 neutrons.

$^{222}$Rn has four valence protons that partially fill an $h_{9/2}$ level and ten valence neutrons, that completely fill the $g_{9/2}$ level. This is four more protons and ten more neutrons than the "double magic" isotope of lead $^{208}$Pb which has closed neutron and proton shells, and is the last double magic isotope in known nuclei. Due to the filling of the $g_{9/2}$ level, $^{222}$Rn is the longest lived isotope for all Rn nuclei. However, due to the nuclear force range coupled with the partial filling of the proton level, in the end $^{222}$Rn has a half-life of only 3.82 days. Furthermore, like many of the nuclei in the valley of stability, $^{222}$Rn decays via alpha emission. Elements that have stable nuclei, with fewer than 84 protons (see the black ridge down the center of Figure 9), typically have isotopes that will decay via beta emissions when too many or too few nucleons are present. Very rarely, and with a significant increase in the number of neutrons needed for stability, some elements will emit an alpha particle. The valley of stability is marked by a region where the typical mode of decay is alpha emission. Thus as an isotope, $^{222}$Rn displays unusual nuclear dynamics compared to most other elements and their isotopes.

In fact, more than half of all known Rn isotopes, 30 out of 40, can decay via alpha emission – for 23 out of 40, alpha decay is the dominant mode, while 17 isotopes decay predominantly by $\beta$- decay and 5 decay predominately by $\beta$+. Only about 12 elements with fewer than 99 protons (elements with 99+ protons are all man-made) have more than half of their isotopes decay via alpha emission. Some of this decay behavior can be explained by models of the nuclear strong force as a combination of a 3-dimensional harmonic oscillator along with a spin-orbit interaction. It should also be noted that Rn, with such short lived isotopes, only "naturally occurs" due to its production during the $^{238 \text{ and } 235}$U and $^{232}$Th decay chains.

## 4.0 Theory & Experimentation:

### 4.1 Theory on GSI

It is tempting to assume that either some standard model physics, or some known type of detector anomaly, can explain the key features of the GSI data. What makes the data so unique, and beyond a standard model or detector glitch explanation, is that the data can be predicted using a vector sum of two low-energy radiation fields. Known phenomena, even accounting for neutrino oscillations and highly biased (highly sensitive) detectors, do not generally respond to directional information for low energy photonic fields.

Consider some potential explanations that do not involve exotic matter couplings including: 1.) Could a known particle, e.g. solar neutrinos, be responsible for the observations, 2.) Is it possible that the detector is hypersensitive to radiation – acting more like night-vision and amplifying low energy radiation fields, 3.) Can levels of radon in the container oscillate within the apparatus, creating a cyclic effect, timed with background radiation heating, 4.) Could electrical oscillations, due to power consumption, create annual and diurnal effects or 5.) Can the data be explained by weather effects as suggested by the work of Pomme et. al. **[13-14]**. Each of these potential explanations would fail to describe two very significant, salient features: 1.) the data drop more rapidly than Gaussian, though the solar irradiance follows a more or less Gaussian distribution centered at noon, and 2.) the data show a peak at approximately 6pm with a duration of approximately 4hours (daily average).

The problem with solar neutrinos as an explanation: Solar neutrinos tend to oscillate as they move through the Earth, leading to a brighter irradiance of electron neutrinos ($\nu_e$'s) at midnight versus what is seen through the day. Solar irradiance in the northern hemisphere drops during the winter due to the Earth's tilt. This means that solar neutrinos will then have to travel through a portion of the Earth to reach a detector in the northern hemisphere. The propagation of neutrinos through the ground would lead to oscillations that increase (not decrease) the numbers of $\nu_e$'s making it into the GSI detector. One should then expect to see an enhancement in radioactive decays depending on $\nu_e$'s at midnight not noon. Furthermore, there is no reason for solar neutrinos, which can pass through the entire mass of the Earth, to rapidly drop off between 3pm and 4pm, only to resurge at 6pm, by as much as 20-30% of the observed noon rates.

If the detector, due to electronic biasing, were to become ultra-sensitive to heat or visible light wavelengths, one would find it difficult to explain how information on the relative direction of solar and albedo radiation fields can be retained. Having albedo radiation around noon directed upwards from the Earth, while solar radiation is directed downwards towards the Earth, would result in an increase of electronic avalanches stimulated by both low energy radiation fields. What is seen in the data, is that the albedo and solar radiations appear to cancel each other around 3pm as the albedo radiation increases. This is not consistent with a detector where the bias leads to ultra-sensitivity to low energy photons. Similarly, there is no mechanism that would drive radon particles out of the detector as it heats up under the influence of solar energy. However, the additional heat energy due to albedo radiation reduces the effect allowing the radon to recover between 4pm and 6pm. Simply stated: *heat flowing in one direction cannot be used to remove heat flowing in another direction.*

Dr. Pomme et. al. raised the possibility of environmental conditions such as weather playing a role in anomalies observed in the GSI data. A review of 2011 data showed no significant correlation between rain (light to heavy) and fluctuations in the rates of gammas detected in the data. In spite of the fact that Israel's rain falls mostly in the late fall through the winter, the nosiest period for the data as a whole, there are no more than statistical overlaps between the periods of even heavy rain and the periods of strong variations from a typical day. Thus the weather does not correlate as suggested, see Figure 10. Additionally, Sturrock et. al (see Reference [5]) looked for correlations between temperature and the GSI signal. Figure 11 shows a sample of both weather and GSI data from September of 2009. Not only does the graph of temperature show a single peak, roughly centered at 2pm local time, there is no characteristic drop between 1pm and 3pm as seen in GSI. Additionally, both data sets shows significant, temporal fluctuations that are not mimicked in the other.

**4.2 Experimentation**

If the GSI observations are due to a Primakoff mechanism, this can be tested experimentally. There have been a significant number of experiments, proposed and executed, attempting to see the Primakoff coupling **[15-27]**. Previous approaches focused on propagating photons, usually utilizing a laser, through an external, table-top magnetic field and searching for selected absorption or evidence of scattering. The Primakoff coupling scales with both the strength of the field (B) and the length of the region (L) through which the photons travel. While table-top magnets can achieve fields up to 52 T ($5.2 \bullet 10^{+5}$ gauss), the distance scales are usually of order 0.05-10 m. For the Earth-Sun system, the Earth has surface magnetic field of ~ 0.5 gauss that stretches for thousands of kilometers ($10^{+6}$ m). In a follow up paper, a comparison between the Earth-Sun ($B \bullet L$) parameters and what can be achieved with a terrestrial, table-top experiment will be presented.

What is also notable, based on the GSI data, is an absence of experiments using a geometry where any collective, focusing effects would be expected. Thus far, experimental approaches were staged to see the impacts of axions on photon beams (cavity experiments) or to cause axions to absorb and re-radiate photonic energy ("light through a wall"). The sheer number of axionic particles encountered in cavity approaches, for example, is limited by the cross sectional area of a laser beam coupled with the length scale for the magnetic field. No attempts to date have been made that would take advantage of a wide area of axionic particles expected near the surface of the Earth. Nor have experiments been performed to take advantage of potential sensitivity of nuclei to minute changes in alpha energy, that a neutral, weakly-interacting particle can uniquely cause.

**5.0 Summary:**

The Geological Survey of Israel (GSI) experiment was designed to detect subtle changes in the radon decay for purposes of investigating behavior of radon in an enclosed environment. To this end, the detection scheme was designed to be particularly sensitive to decay products of radon. Details of the experimental apparatus can be found in earlier analysis by Steinitz, Piatibratova and Sturrock **[1-4]**. The system was sealed for the entire 10 years of data taking and there are no internal, movable parts. The datalogger and detectors were all powered via battery – isolating them from the local power grid, thus preventing influence from human power consumption. The data over six years exhibit a consistent diurnal and annual variation. The annual oscillations, which follow solar irradiance, show that low energy radiation correlates with the amplitude and time of the primary peak. Furthermore, the sudden drop at 3pm, followed by a second peak at 6pm can be created by vector addition of solar and albedo radiation, which is a process that has not been predicted for radioactive nuclear decay.

The above observations appear to be consistent with a mechanism such as the Primakoff, effect whereby photons couple to both magnetic fields and ambient, axionic matter. In such a model, two radiation fields influence a background of axionic dark matter in such a way as to focus or change the trajectory of these particles. Additionally, this allows two radiation fields in the presence of a magnetic field, in a Primakoff model, to compete with each other and produce a rapid drop off as observed between noon and 3pm. Thus, vector behavior of photonic radiation can be observed in nuclear radiation. What remains to be done is staging an experiment that takes advantage of both nuclear sensitivity through alpha decay as well as the possibility of focusing of streams of dark matter axionic particles onto a target.

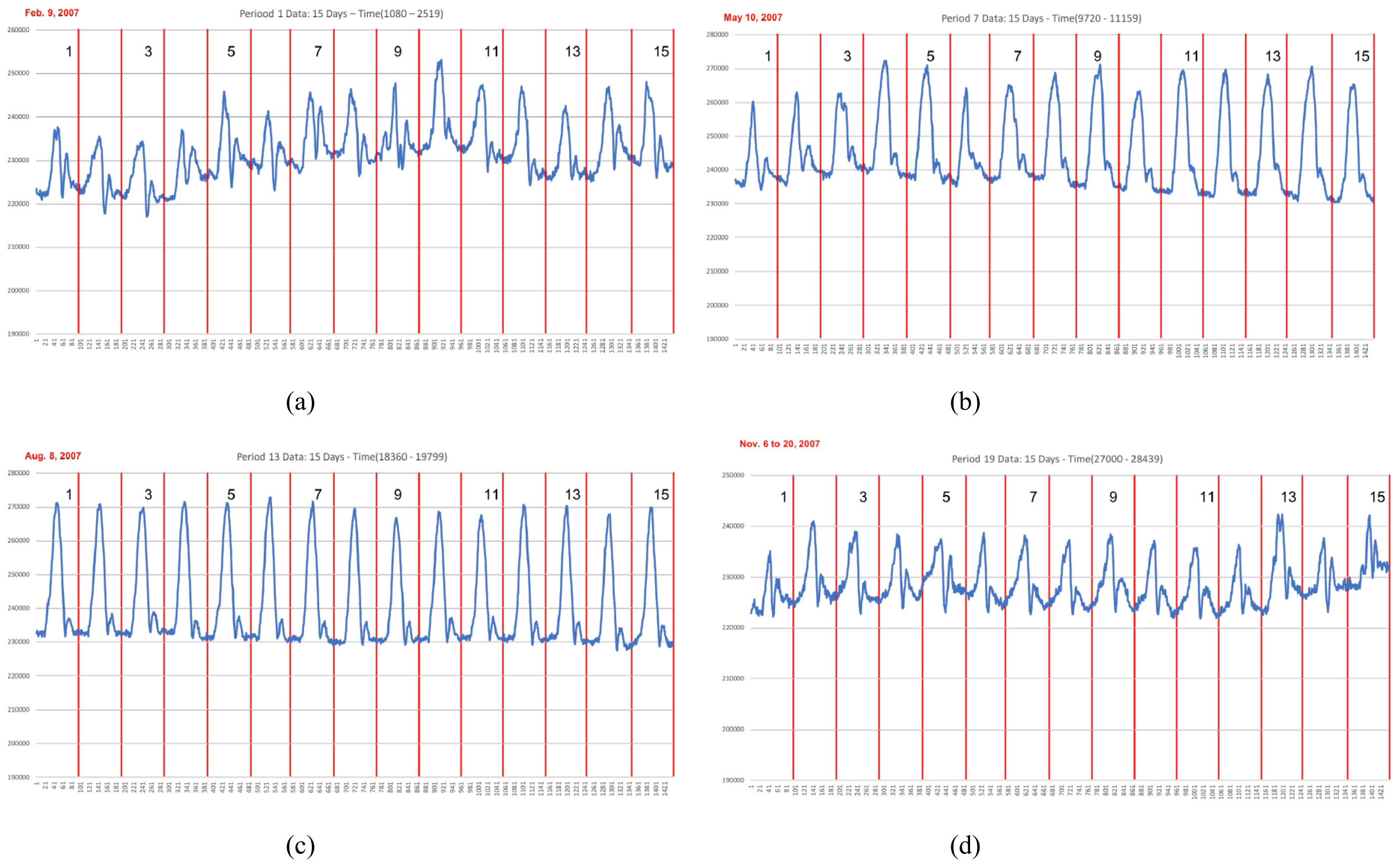

Figure 1: Sample data for: (a) Winter 2007, (b) Spring 2007, (c) Summer 2007 and (d) Fall 2007

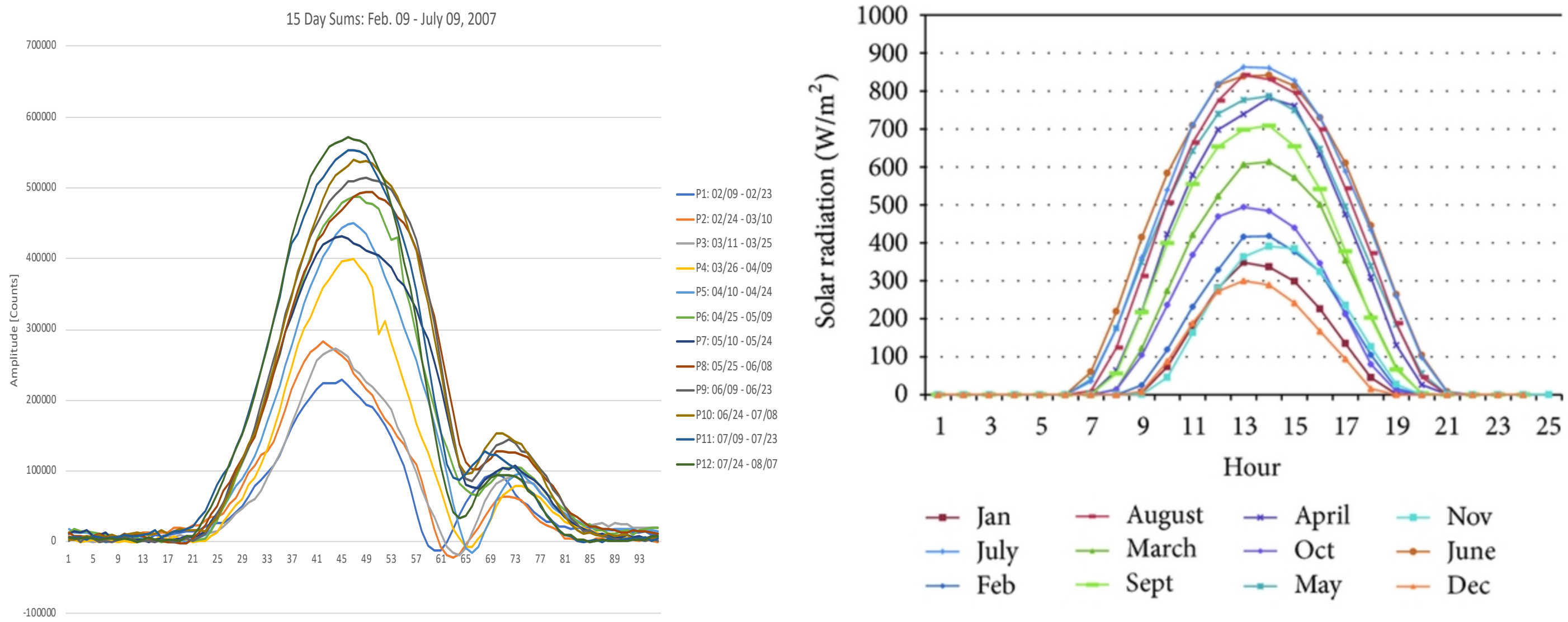

Figure 2: (Left) Sum of 15 days into single plots and (Right) Solar Irradiance over one year period

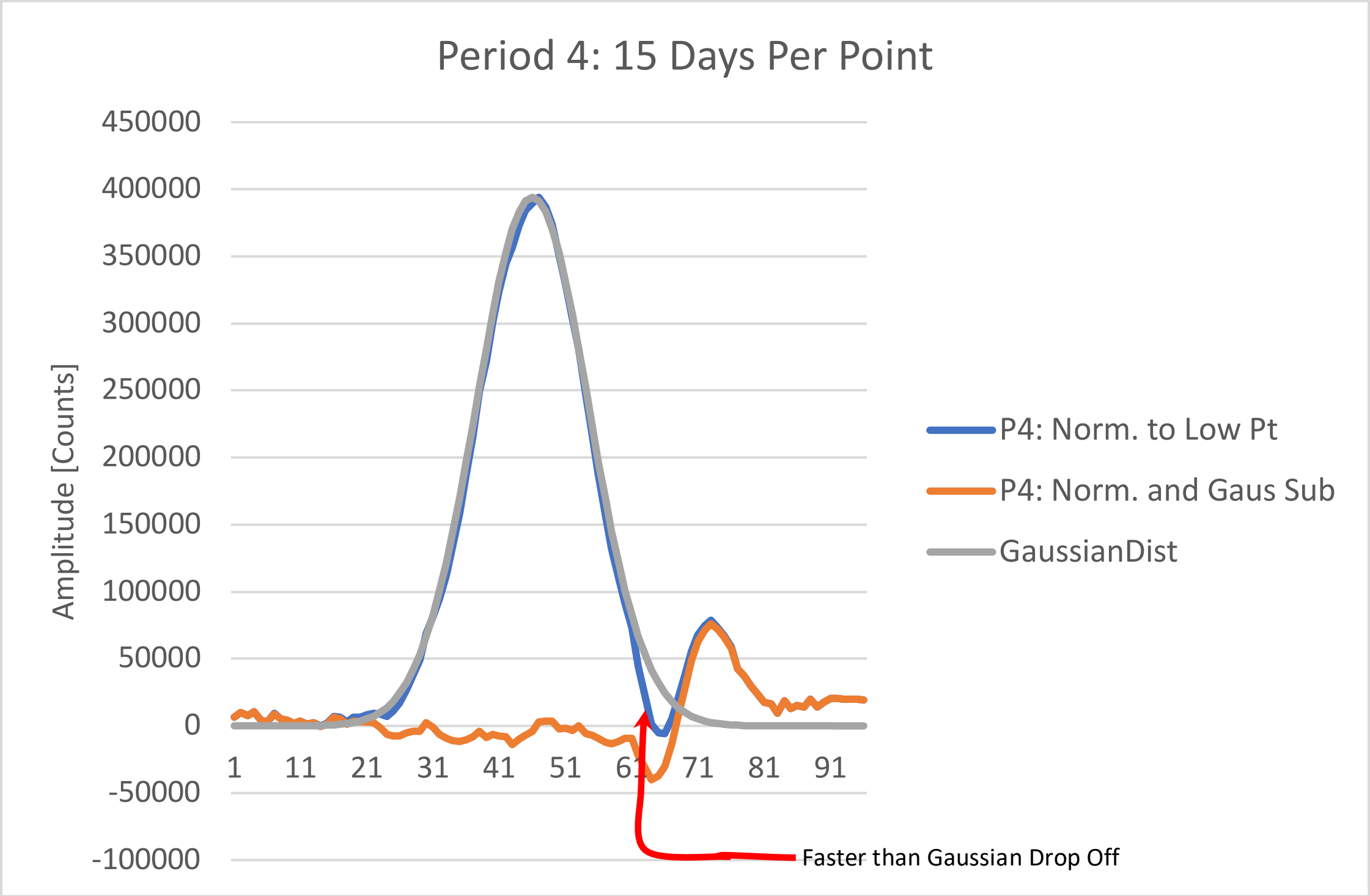

Figure 3: Sample of GSI data with a Gaussian function overlaid.

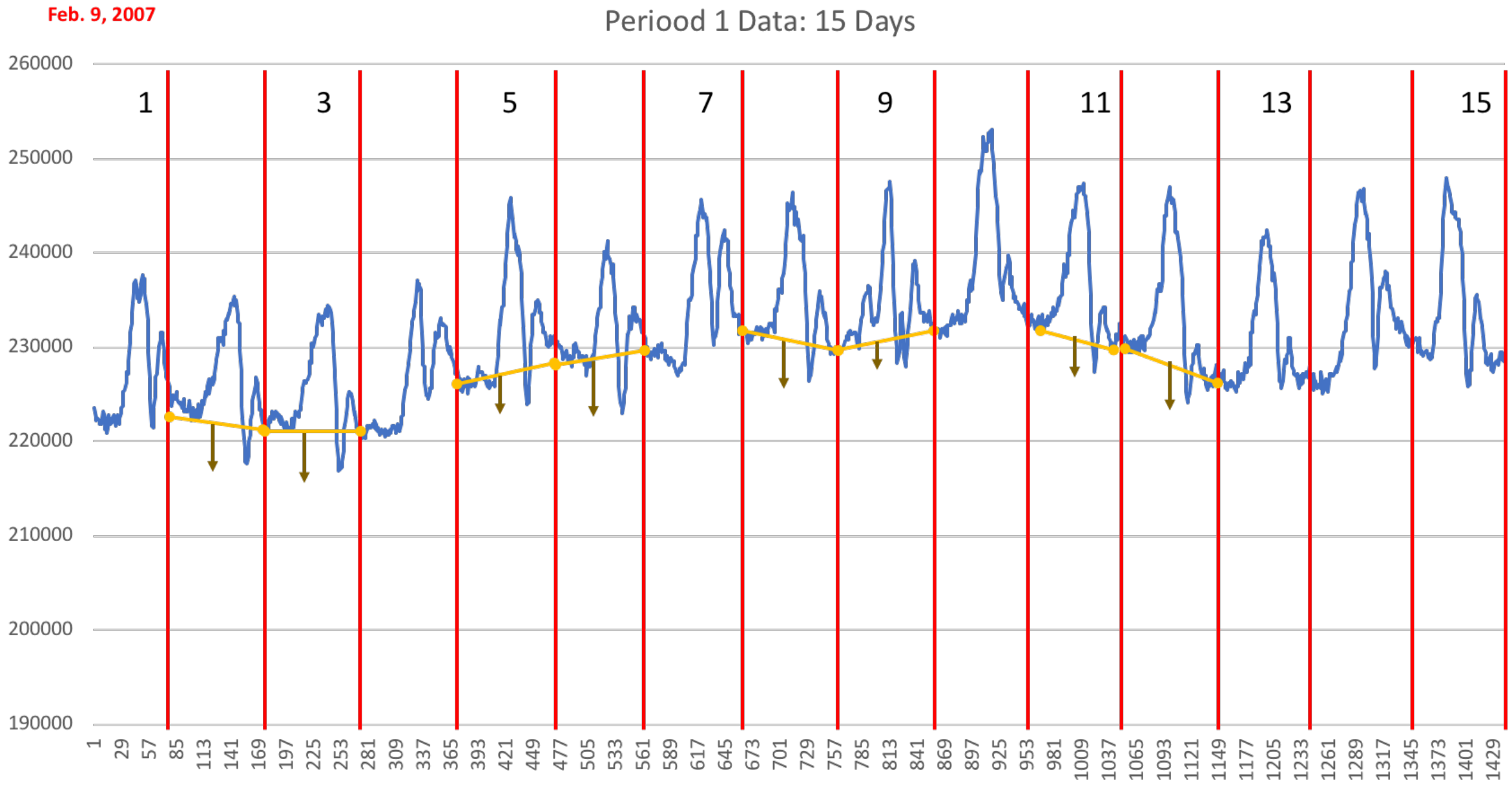

Figure 4: Data from February 2007 showing how detection rates drop near 4pm daily; a baseline shown in yellow is defined by peak sidebands regions 12am and 1am (next morning); arrows show how rates drop below the defined baseline on many days.

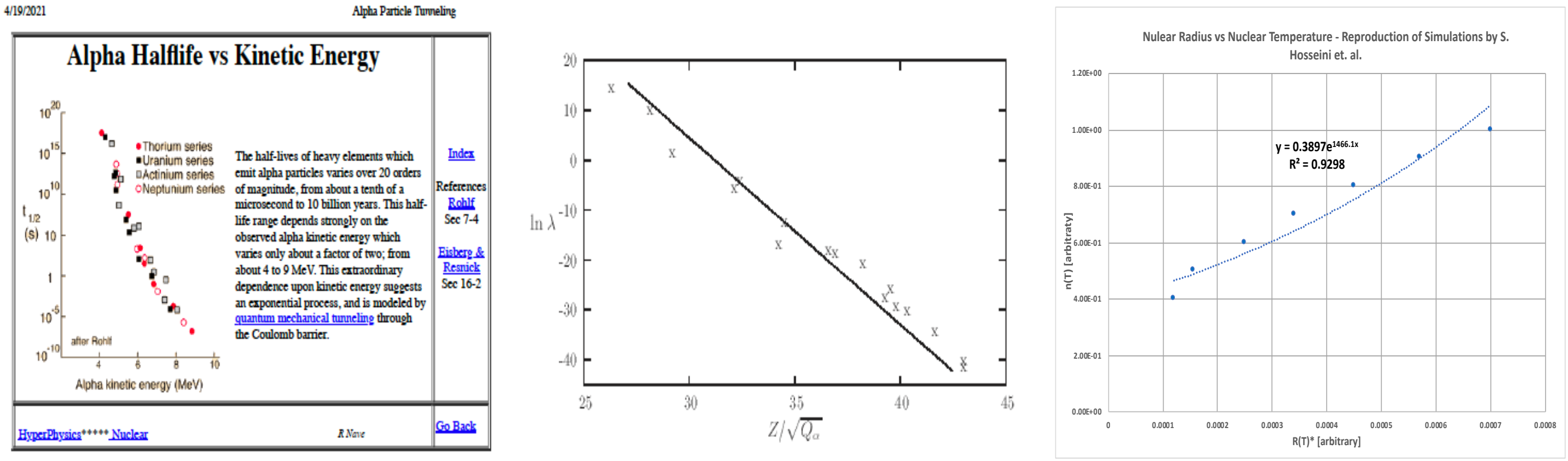

( a ) ( b ) ( c )

Figure 6: (a) Graph of data showing alpha decays for several nuclei as a function of Energy from HyperPhysics link (b) Graph of data showing alpha decay as a function of the Gamow (G) factor taken from undergraduate textbook, University of South Hampton, UK and (c) From S. Hoosseini' work on nuclear radius and temperature,

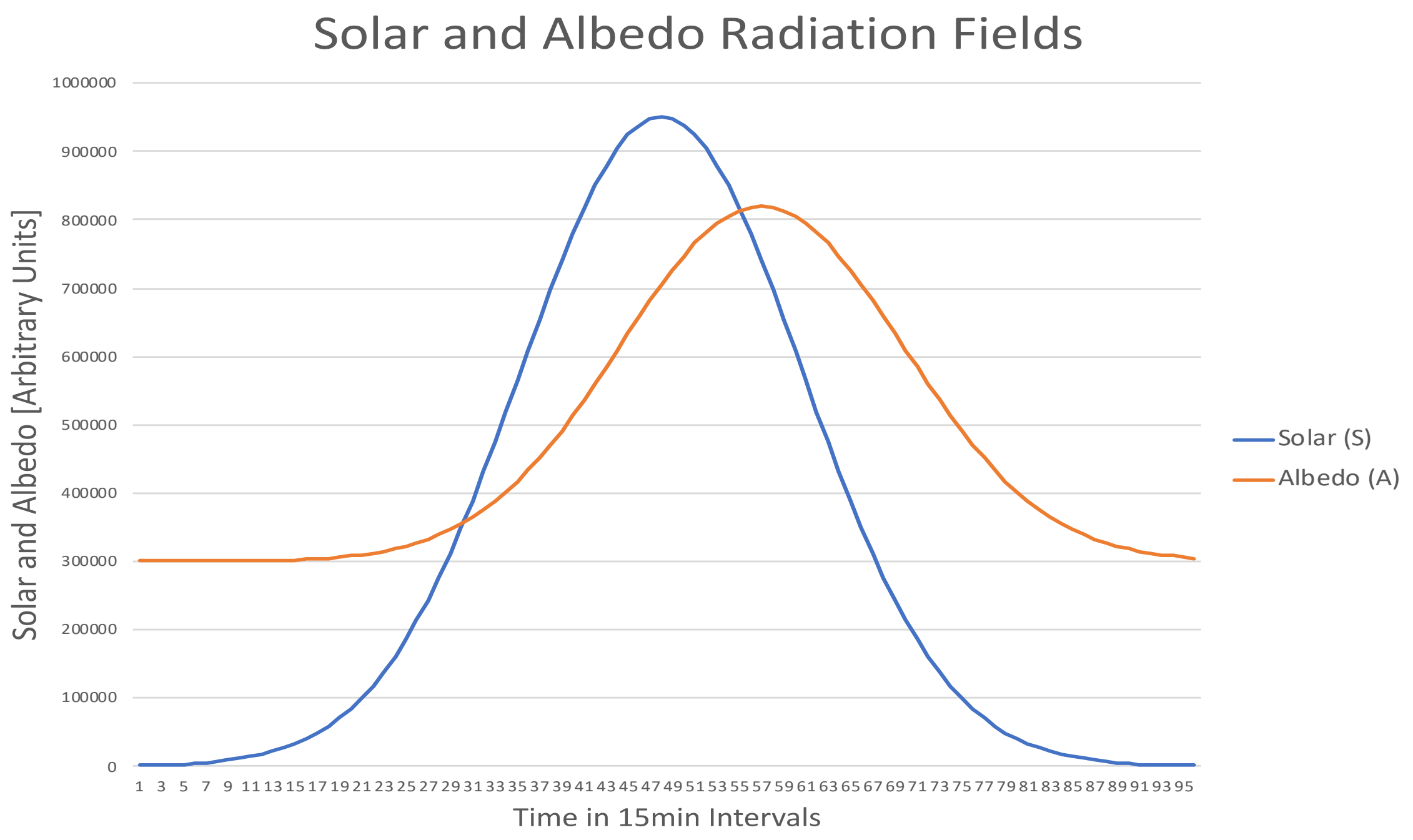

Figure 7: Solar and Albedo radiation near the Earth's surface.

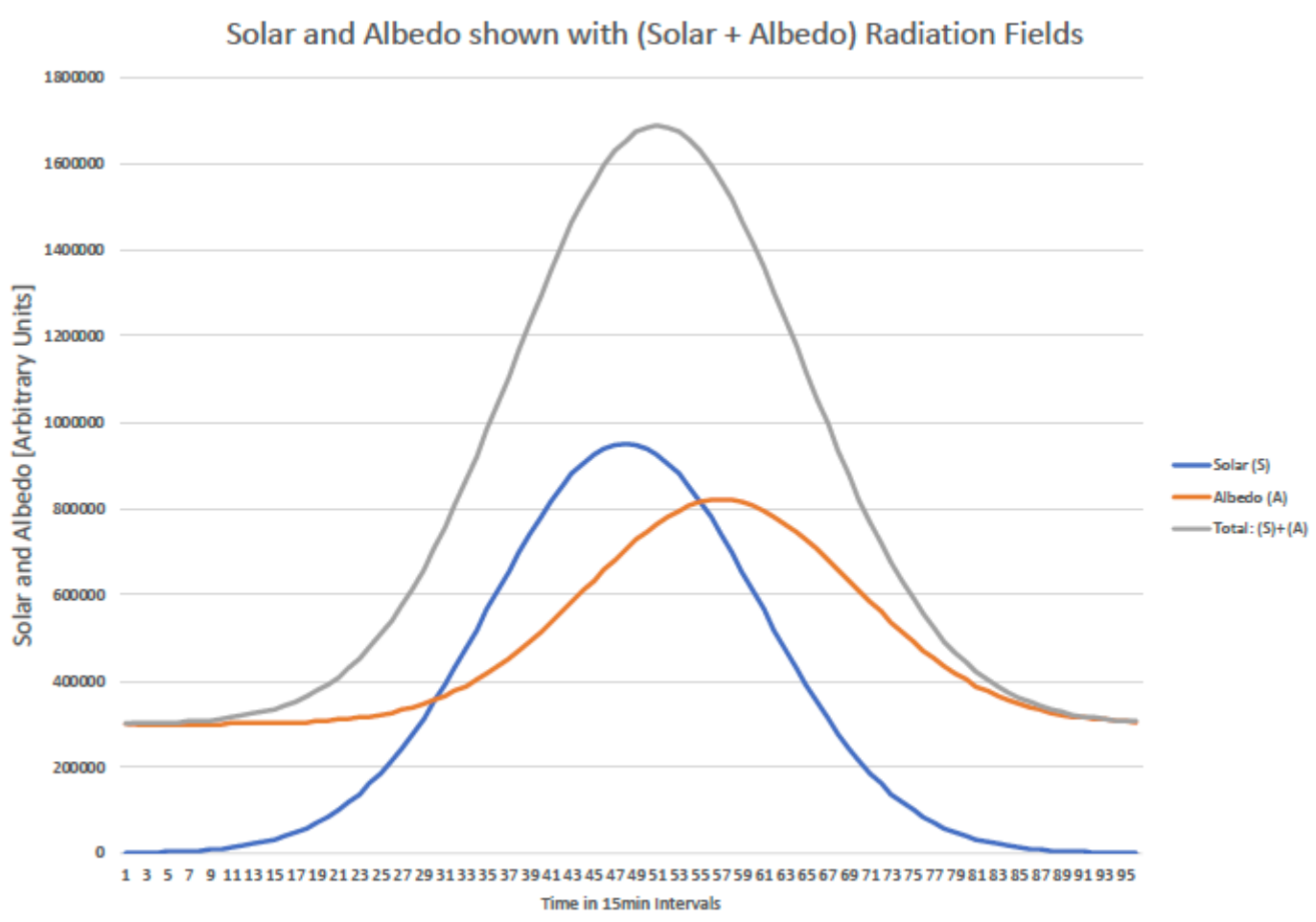

(a)

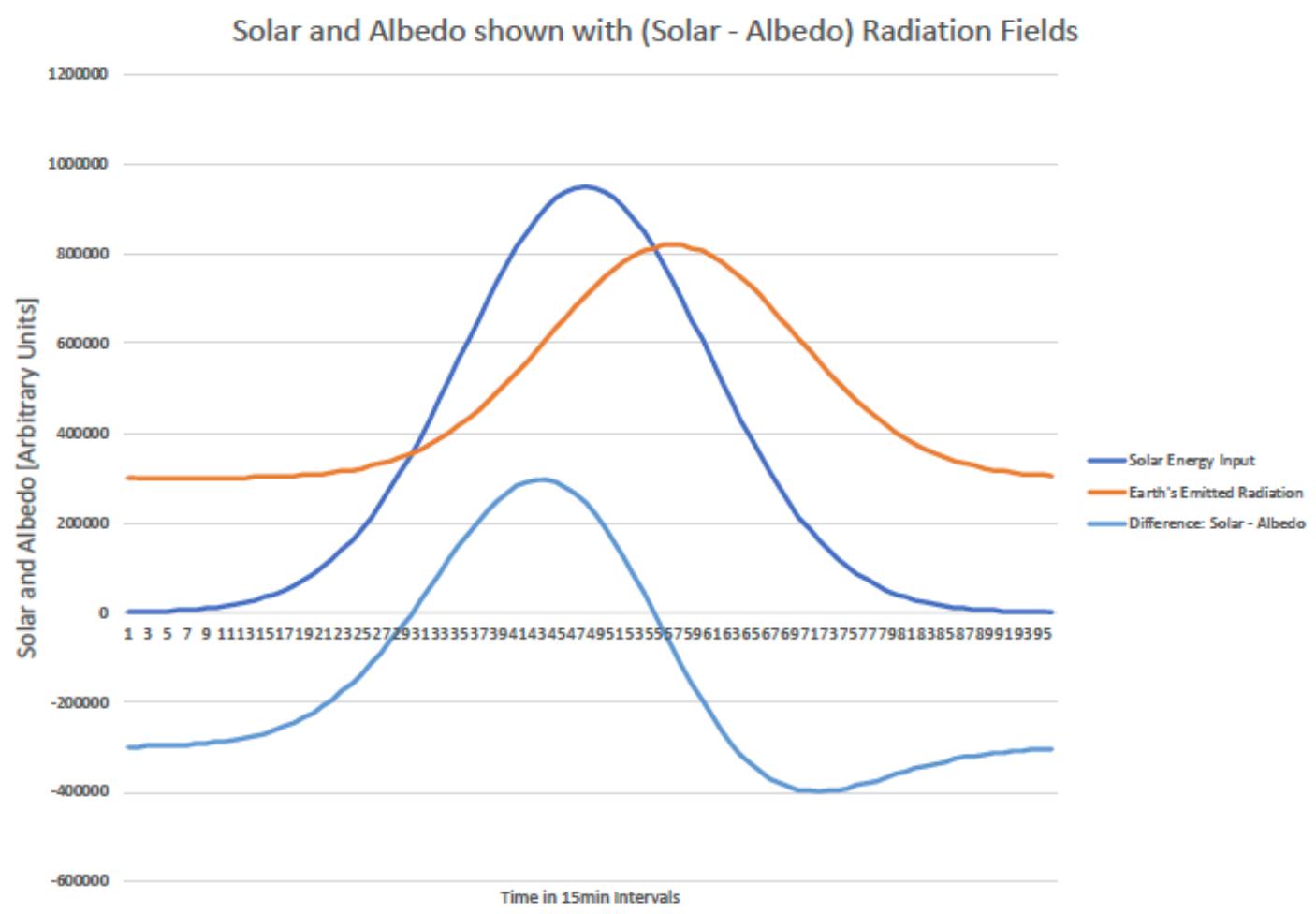

(b)

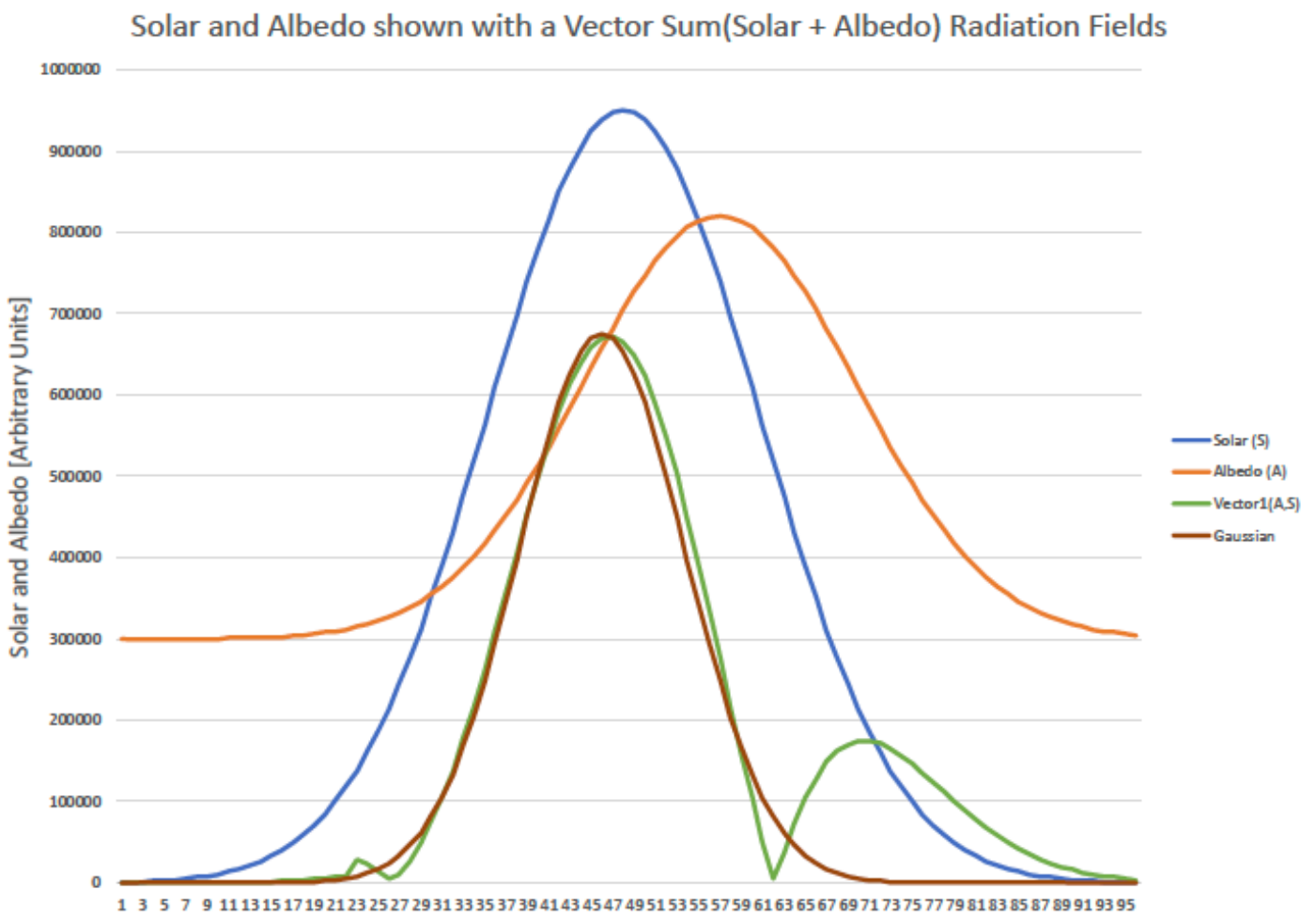

(c)

Figure 8: (a) (gray) Total radiation observed by summing solar and albedo for each 15min period, (b) (light blue) Difference between solar and albedo radiation for each 15min period and (c) (green) A vector sum of solar and albedo radiation, taking into account the relative directions of each other over a day.

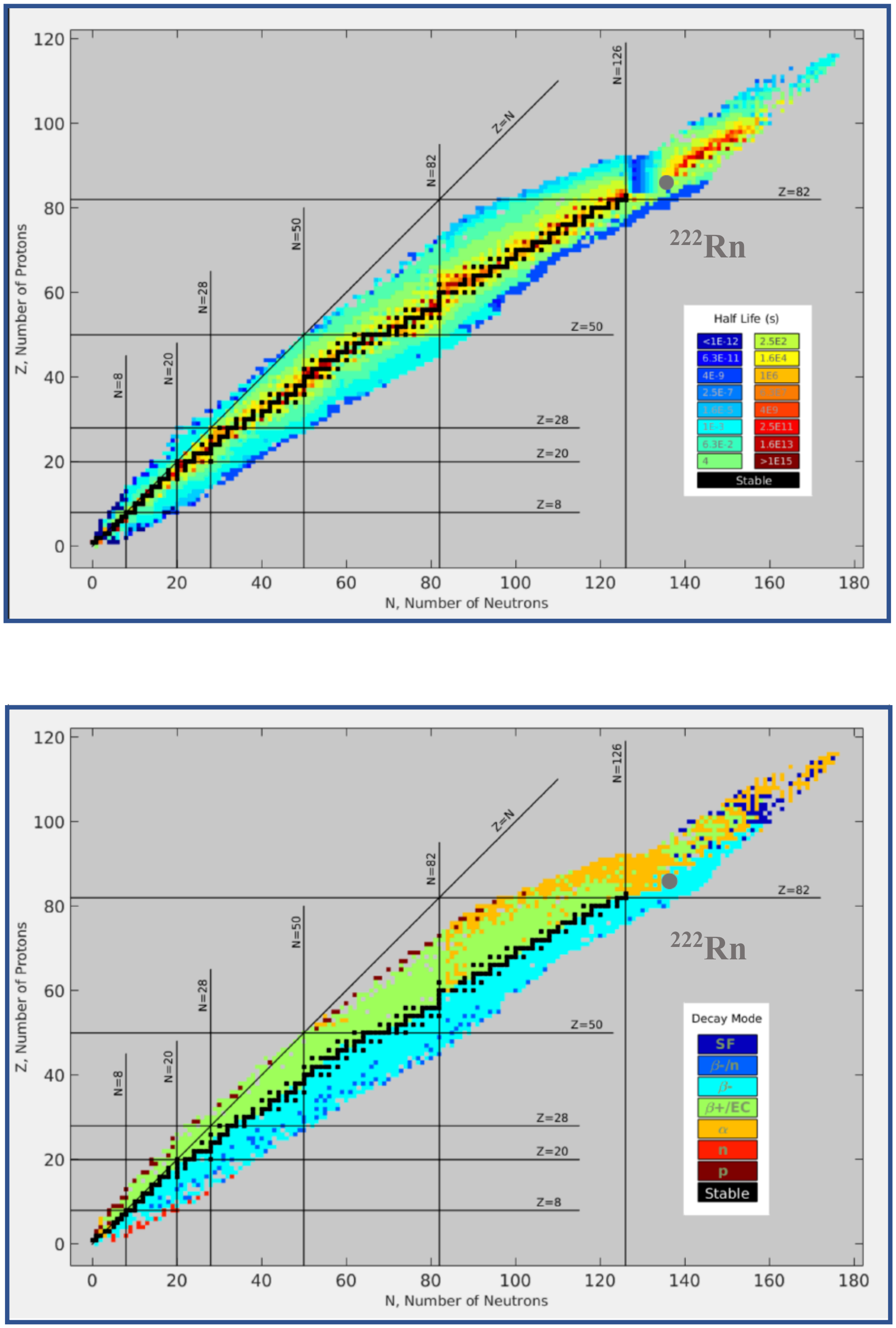

Figure 9: (Top) Half-life for all measured nuclear isotopes with $^{222}$Rn denoted by a gray circle and (Bottom) Decay modes for all measured nuclear isotopes

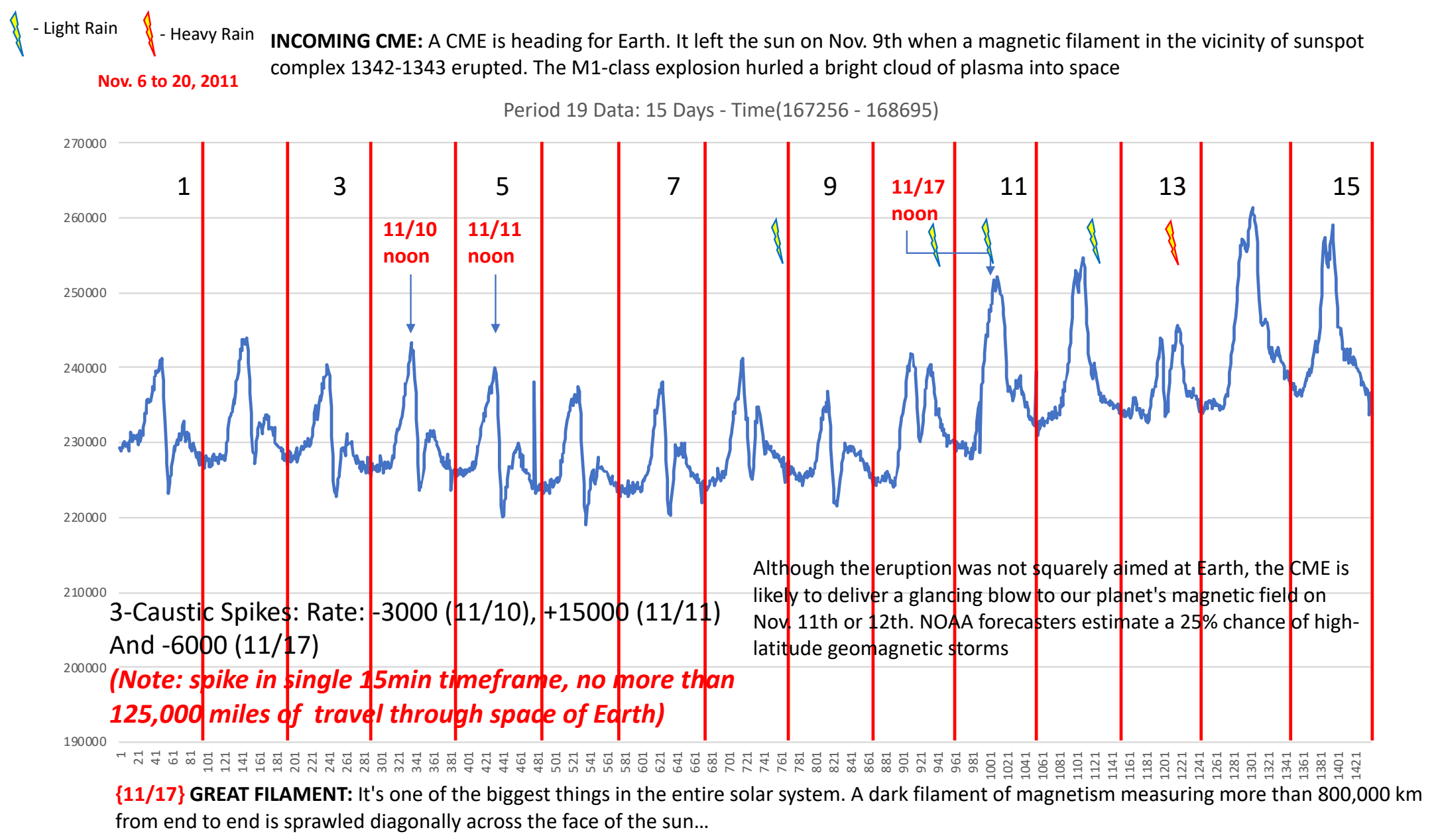

Figure 10: A look at weather conditions in Jerusalem, Israel in November of 2011; rain report has been divided into four-six hour periods (12am-6am, 6am-12pm, 12pm-6pm and 6pm-midnight); of the five periods of rain fall, one of which is very heavy, none of the observed fluctuations in November coincide with a period of rain; note there are sixty 4-hour periods in the graph and approximately 8 instances of abnormal behavior.

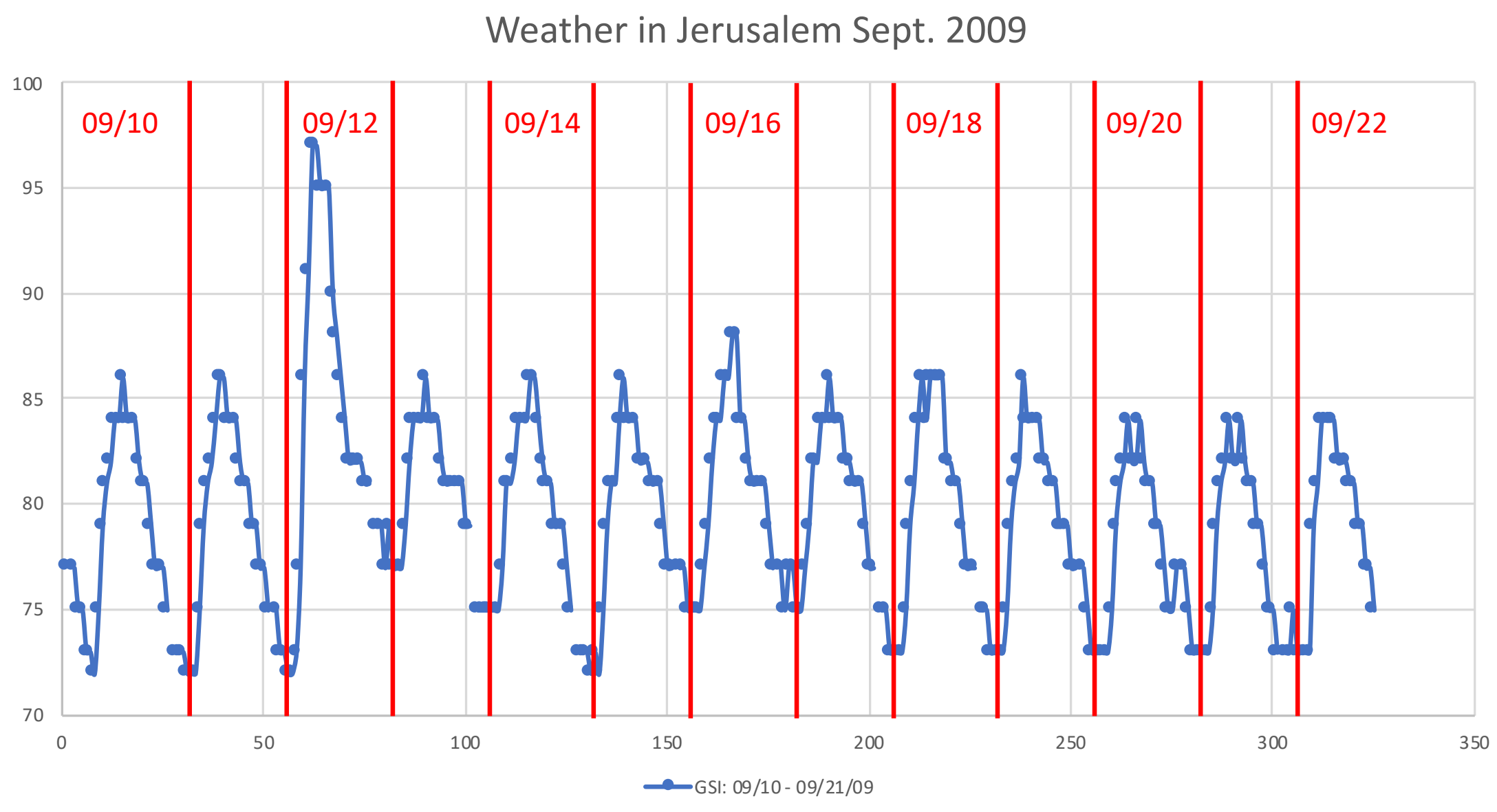

(a)

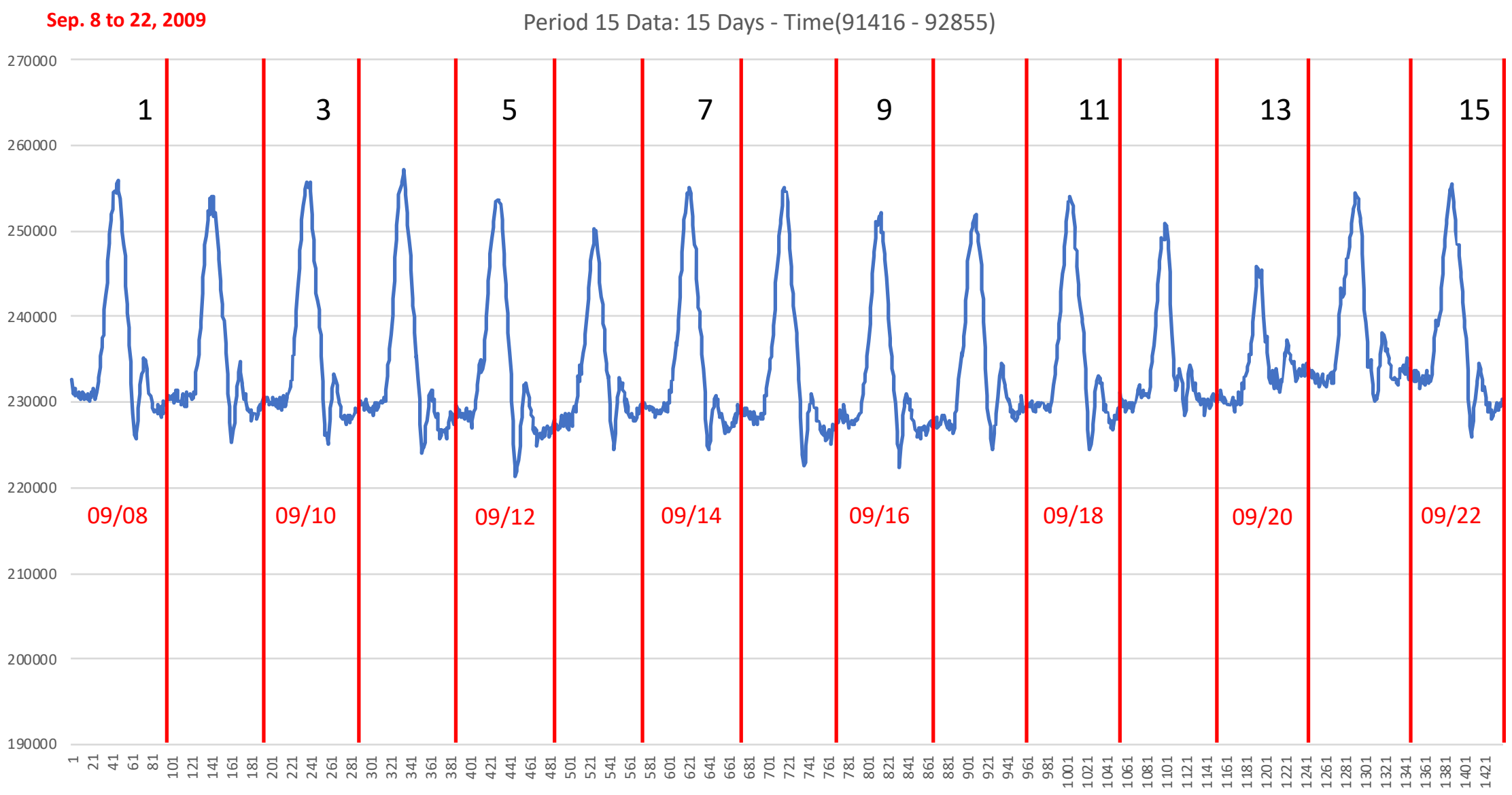

(b)

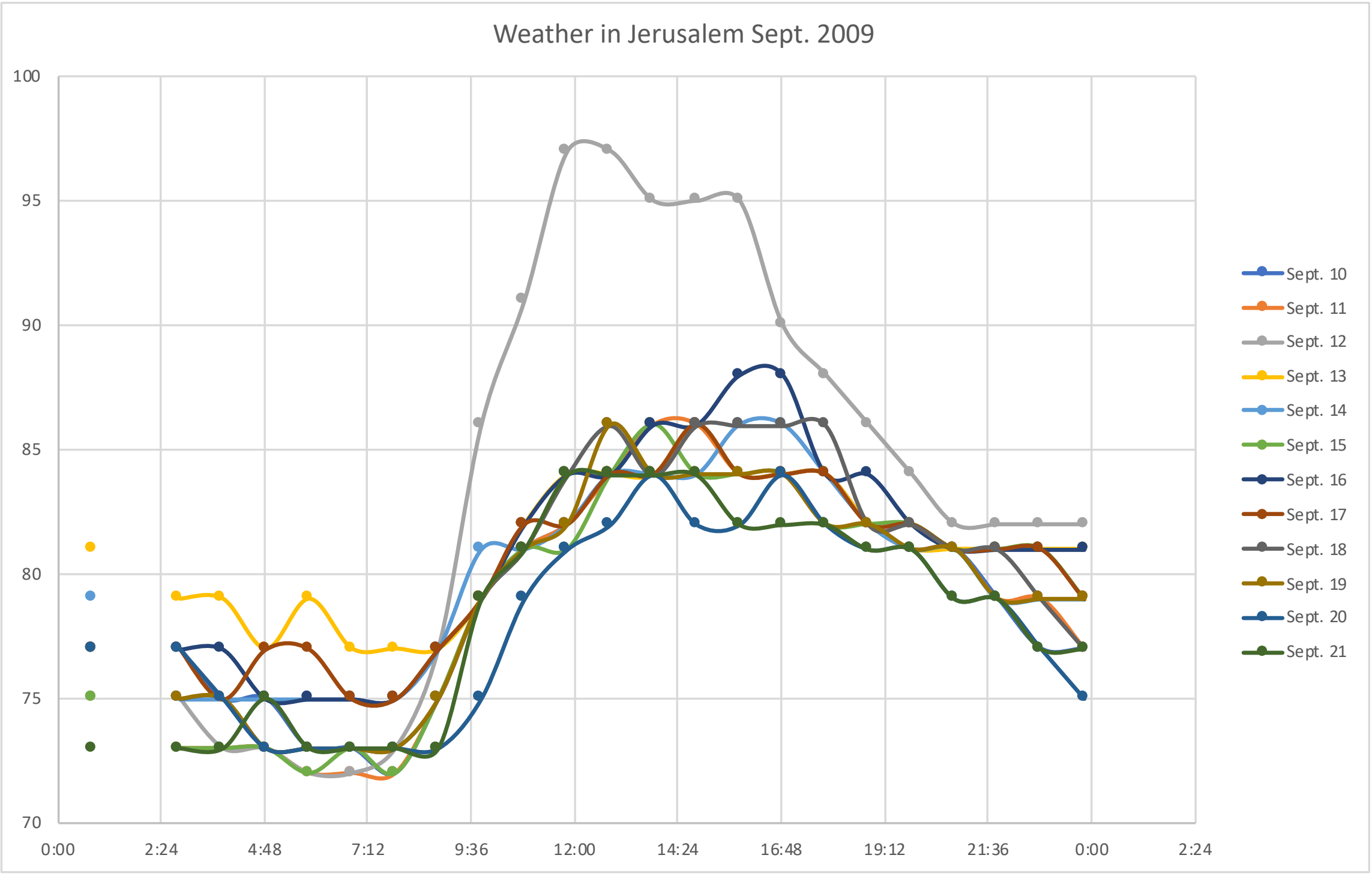

(c)

Figure 11: (a) Weather in Jerusalem between 09/10 – 09/22/09, shown as a day-by-day for comparison to GSI; (b) GSI data sample, showing September 8$^{th}$ through 22$^{nd}$, note the distinct features of the data over a 24hr period including ; (c) Weather data for Jerusalem between 09/10 and 09/21/09 showing days overlapped for amplitude comparison